\documentclass[runningheads]{llncs}
\usepackage[T1]{fontenc}
\usepackage{graphicx}
\usepackage{times}
\usepackage{soul}
\usepackage{url}
\usepackage[hidelinks]{hyperref}
\usepackage[utf8]{inputenc}
\usepackage[small]{caption}
\usepackage{graphicx}
\usepackage{amsmath}
\usepackage{booktabs}
\usepackage{algorithm}
\usepackage{algorithmic}
\usepackage[switch]{lineno}
\usepackage{amsmath,amsfonts,amssymb}
\usepackage{algorithmic}
\usepackage{algorithm}
\usepackage{array}
\usepackage{textcomp}
\usepackage{stfloats}
\usepackage{url}
\usepackage{verbatim}
\usepackage{graphicx}

\usepackage{wrapfig}
\usepackage[hypcap=false]{caption}
\usepackage{amsmath}
\usepackage[capitalise, noabbrev]{cleveref}
\usepackage[font=small,skip=5pt]{caption}
\usepackage{multirow}
\usepackage{tikz,subcaption}
\begin{document}
\title{A Wearable Multi-Modal Edge-Computing System for Real-Time Kitchen Activity Recognition}
%
%
\author{
Mengxi Liu\inst{1} \and
Sungho Suh\inst{1,2} \and
Juan Felipe Vargas\inst{1} \and 
Bo Zhou \inst{1,2} \and
Agnes Grünerbl \inst{1} \and
Paul Lukowicz \inst{1,2}}
\authorrunning{M.Liu et al.}
%
\institute{German Research Center for Artificial Intelligence (DFKI), Kaiserslautern, Germany \and
Department of Computer Science, RPTU Kaiserslautern-Landau, Kaiserslautern, Germany
\email{firstname.lastname@dfki.de}}
\maketitle              
\begin{abstract}
In the human activity recognition research area, prior studies predominantly concentrate on leveraging advanced algorithms on public datasets to enhance recognition performance, little attention has been paid to executing real-time kitchen activity recognition on energy-efficient, cost-effective edge devices. Besides, the prevalent approach of segregating data collection and context extraction across different devices escalates power usage, latency, and user privacy risks, impeding widespread adoption. This work presents a multi-modal wearable edge computing system for human activity recognition in real-time. Integrating six different sensors, ranging from inertial measurement units (IMUs) to thermal cameras, and two different microcontrollers, this system achieves end-to-end activity recognition, from data capture to context extraction, locally. Evaluation in an unmodified realistic kitchen validates its efficacy in recognizing fifteen activities, including a null class. Employing a compact machine learning model (184.5 kbytes) yields an average accuracy of 87.83 \%, with model inference completed in 25.26 ms on the microcontroller. Comparative analysis with alternative microcontrollers showcases power consumption and inference speed performance, demonstrating the proposed system's viability.

\end{abstract}
\section{Introduction}

Human activity recognition (HAR) provides a promising method for personal health monitoring and is exploited in many daily scenarios like healthcare centers, sports monitoring, and smart home, and it has emerged as a paramount study direction for the pervasive community \cite{kumrai2020human}. 
Although researchers have explored many application scenarios of HAR, kitchen activity recognition has not yet been widely studied in comparison with others. 
A kitchen is an important place that people visit almost every day. 
Human activities in the kitchen are more closely related to their eating habits than in other environments. For example, opening microwave ovens and refrigerators may reflect a person's long-term food consumption frequency. 
Unhealthy dietary habits such as overeating, eating too frequently, or at inappropriate times can lead to many diseases.
Therefore, continuous monitoring of kitchen activity can help people build their model of dietary habits and encourage them to develop healthier dietary habits. 
At the same time, the physician can also benefit from the dietary data to diagnose possible diseases.  

Many existing works have focused on extracting activity-related context using complex algorithms on public datasets \cite{yang2016super,du2015hierarchical,yang2012recognizing} or developed new sensor modalities \cite{bian2022human} to improve recognition accuracy, while few of them studies implemented real-time kitchen activity recognition on a low-power, low-cost edge computing device. 
Deep learning (DL) is the primary algorithm used in state-of-the-art research for HAR.  
DL models with millions of parameters bring challenges to the deployment in resource-constraint microcontrollers, resulting in data processing on the cloud, which requires additional data transmission causing drawbacks during the application in real life. For example, the risk of user privacy exposure, unexpected power consumption, and unnecessary latency.
The edge-devices-centric architectures can be more power-efficient, provide better privacy, and reduce latency for inference \cite{zaidi2022unlocking}.
With the help of the tiny machine learning framework, integrated sensors in the smartphone (inertial measurement unit (IMU), camera, barometer, and microphone) have turned the smartphone into a very competitive multi-sensor edge-computing platform \cite{milenkoski2018real}. 
However, limited sensor modalities make smartphones unsuitable for complex activity recognition in many applications, for example, food preparation activity in kitchens, which can be addressed by the smartphone camera though, privacy issues prevent it from wide utilization. 

To address the problems mentioned above, in this work, we present a wearable multi-sensor edge-computing platform for real-time human activity recognition in the kitchen.
The hardware contains six sensors, one low-power consumption microcontroller, and one high-performance microcontroller. 
Using two microcontroller units (MCUs) improves the system's data processing performance and extends multiple sensors' peripheral interface for data acquisition. Users can deploy the model to different microcontrollers to obtain high computational performance or energy efficiency.
Besides, we deployed different machine learning models based on multi-channel time-series convolutional neural networks (MC-CNN) \cite{yang2015deep} and DeepConvLSTM \cite{ordonez2016deep} for kitchen activity recognition on these MCUs to study the hardware performance. 
The model inference process is implemented locally, by which the raw data from users can be well protected locally, and the power consumption of data transmitted to the cloud can be avoided. 

Overall, in this paper, we present the following contributions:

\begin{itemize}
    \item We developed a wearable multi-sensor edge-computing system to recognize fourteen human activities in a kitchen scenario, which consists of six sensors and two MCUs. 
    \item We deployed two kinds of popular neural network models for HAR on our proposed hardware and two other microcontrollers and compared their performance.
    \item We demonstrated the performance in detecting fifteen activities, including a null class, by an 184.5-KByte neural network model with an average accuracy of 87.83\% and an average inference time of 25.26 ms.
\end{itemize}


\section{Related Works}
\label{sec:relatedworks}


HAR benefits many areas, like healthcare monitoring, fitness tracking, and ambient-assisted living. Researchers have explored many methods to acquire data related to human activity. 
These methods can be divided into two groups: computer vision-based methods and sensor-based methods.
Computer vision-based methods have achieved high recognition accuracy in many application cases by using deep convolutional neural network (CNN) \cite{yang2016super,du2015hierarchical,yang2012recognizing}. 
The sensor-based method is unsubstitutable in some special scenarios like physiological features detection \cite{zhang2020necksense,bharti2018human,gravina2019emotion,mehrang2017human}. 
Similar to our work, HAR in kitchen scenarios based on cameras or other sensors has been explored over the past years. 
For example, Bansal et al. \cite{bansal2013kitchen} used a dynamic SVM-HMM hybrid model to predict nine cooking activities from video information. 
Lei et al. \cite{lei2012fine} proposed a study for fine-grained recognition of kitchen activities using RGB-D (Kinect-style) cameras. 
The proposed system can robustly track and accurately recognize detailed steps through cooking activities.
Luo et al. \cite{luo2019kitchen} demonstrated a minimal and non-intrusive, low-power, low-cost radar-based sensing network system recognizing 15 kinds of activities. 
The related work contains solutions with remarkable recognition accuracy in recognizing kitchen activity and in complex kitchen environments. 
Another similar work for 15 kitchen activities recognition was presented in the work \cite{liu2022smart}. 
However, none of them have evaluated their model on edge devices in real-time, which prevents their widespread use in daily life.

Recently, researchers have studied the deployment of the machine learning model on embedded devices and the execution of real-time inference, which is crucial for a truly pervasive solution, thus bridging the gap between HAR research and commercial products. 
Edge devices are often limited by memory size and computational capacity, which implies CPU speed constraints, and some do not support floating-point operations. 
As a result, the machine learning model running on a workstation or cloud usually cannot be executed directly on the edge device.
To address these problems, many studies have proposed software frameworks and tools for tiny machine learning, 
such as TensorFlow Lite Micro \cite{david2021TensorFlow}, MicroTVM \cite{chen2018tvm}, CMix-NN \cite{capotondi2020cmix}, CMSIS-NN \cite{lai2018cmsis} and STM X-Cube-AI \cite{falbo2019analyzing}. 
In addition, many model size compression methods are also proposed, like pruning \cite{banik2018machine} and conversion/quantization \cite{pourghasemi2020application}. 
With the help of tiny machine learning frameworks, machine learning models running on edge devices have become a reality. 
For example, Wan et al. \cite{wan2020deep} proposed DL models for real-time HAR with smartphones. 
Martin et al. \cite{milenkoski2018real} developed a lightweight algorithm for human activity detection based on Long short-term memory (LSTM) networks. 
The usability of the proposed algorithm is evaluated on real smartphone applications.
Preetam et al. \cite{anbukarasu2022tiny} proposed a machine learning pipeline to extract heart rate from pressure sensor data acquired on low-power edge devices, the size of their designed model was less than 40 kB, which was deployed on an ESP32 edge device.  
Bian et al. \cite{bian2021capacitive} demonstrated real-time hand gesture recognition based on capacitive sensing modality using tiny machine learning and deployed it on the Arduino Nano Sense platform. 
All these works have demonstrated the feasibility of HAR on edge-computing devices. 
However, most of the proposed tiny machine-learning models can only recognize a single category of human activities using multiple input channels from a single sensor modality.

Multi-modal wearable devices may offer advantages over uni-modal wearable devices for HAR, such as higher recognition accuracy, higher robustness (assuming uncorrelated error sources), and broader application scenarios. 
Thus, many works have explored multi-modal hardware.
For example, Zhang et al. \cite{zhang2020necksense} designed a necklace with multiple embedded sensors with four types of sensors to detect eating-related activities.
Bharti et al. \cite{bharti2018human} proposed a multi-modal and multi-positional system called "HuMan". 
The authors used five types of sensors (inertial and environmental) to recognize 21 complex activities in the home, with results up to 95\%. 
Gravina et al. \cite{gravina2019emotion} presented a system based on body-worn inertial sensors combined with a pressure sensor to monitor in-seat activities, by which four ordinary basic emotion-relevant activities were recognized with high accuracy. 

For all the above and to exploit the advantages of a multi-modal design, in this work, we propose a wearable multi-modal edge-computing system for real-time kitchen activity recognition locally.

\section{Hardware Design and Implemented Neural Networks}
\label{sec:hardwaredesign}
\begin{figure*}[!t]
\footnotesize
\centering
\includegraphics[width=0.8\linewidth]{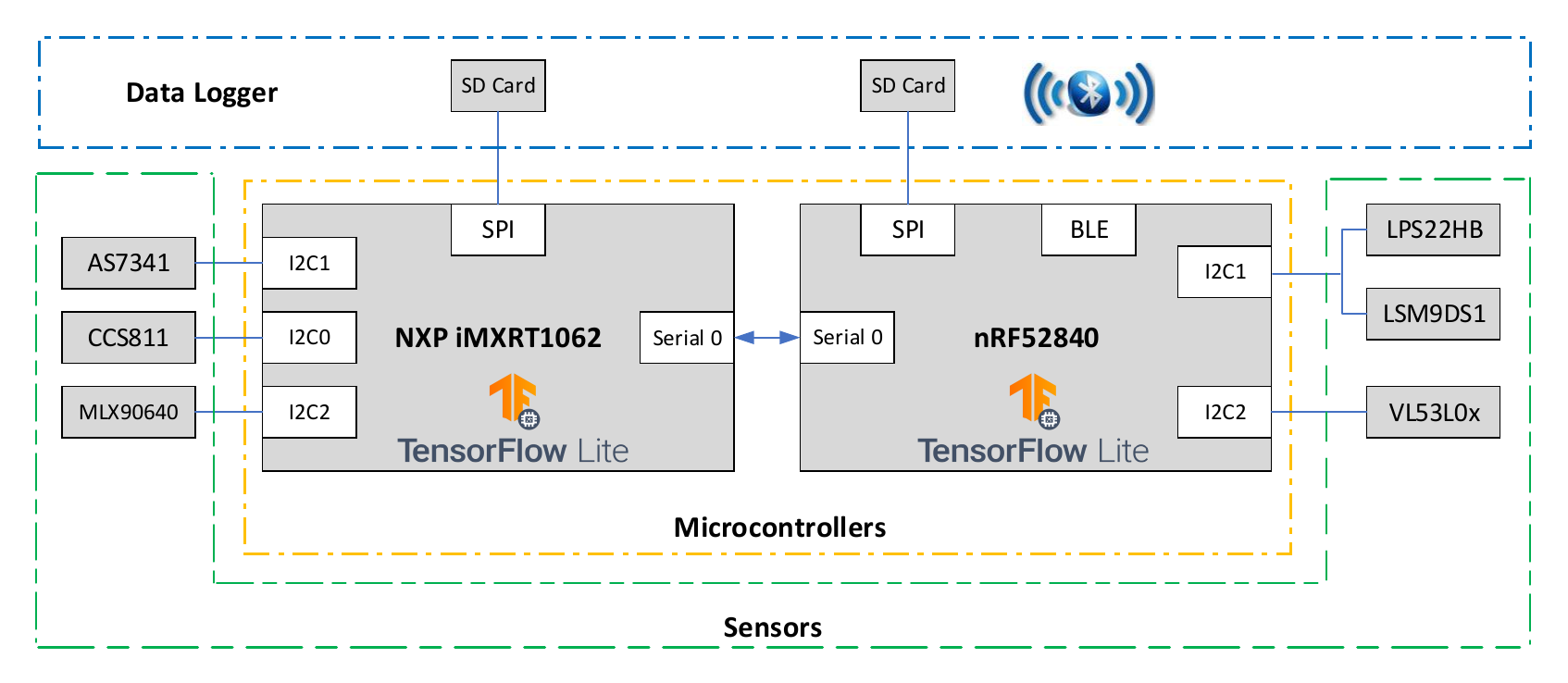}
\caption{Hardware Design of Wearable Multi-Modal Edge-Computing System for Real-Time Kitchen Activity Recognition}
\label{fig: hardware design}
\end{figure*}

As shown in \cref{fig: hardware design}, the proposed wearable multi-modal edge-computing system consists of three main components; the MCU module, the sensor units, and the data logging module. 
All sensors are connected to two MCUs via the Inter-Integrated Circuit (I2C) interface. 
The storage capacity of each MCU is extended through the Serial Peripheral Interface (SPI) connected to an SD card. 
Data transmission between the two MCUs uses the universal asynchronous receiver/transmitter (UART) protocol.
All components are integrated on a prototype PCB board with an overall dimension of 56×64 mm.

\subsection{Sensors}
\begin{table*}[!t]
\centering
\renewcommand{\arraystretch}{1.0}
\footnotesize
\caption{Sensor List}
\label{tab:sensor_list}
\begin{tabular}{c c c l}
\hline
Sensors &Description& Data Channels & Typical Application\\
\hline
AS7431& Optical sensor & 10 & Food and beverages monitor \cite{huang2008near}\\
CCS811& Digital gas sensor  & 2 & Electronic nose \cite{arroyo2020electronic}\\
MLX90640 &  Thermal IR array & 768 & Object detection \cite{naser2020human}\\

LPS22HB &  Air pressure sensor & 1  & Vertical movement \cite{vanini2016using}\\
LSM9DS1& IMU (accelerometer, gyroscope, magnetic meter) & 9 & Fitness monitoring \cite{crema2017imu}\\
VL53L0X& Time-of-Flight ranging sensor & 1& Environment recognition \cite{lakovic2019application}\\
\hline
\end{tabular}
\end{table*}

\cref{tab:sensor_list} lists six sensors available on the proposed edge-computing system. 
As the sampling rate of each sensor varies, these six sensors are divided into two groups and connected to two MCUs, separately. 
A low sampling rate could lead to degradation of measurement accuracy, although it can also provide advantages such as low power consumption. 
The sampling rate range of all sensors is from 3 Hz to 12 Hz, and the final sampling rate of 6 Hz after synchronization is selected in this work.

\subsection{Microcontrollers (MCUs)}
\begin{table*}[!t]
\centering
\renewcommand{\arraystretch}{1.0}
\footnotesize
\caption{Important Parameters of Microcontrollers}
\label{tab: microcontroller}
\begin{tabular}{c c c c c }
\hline
Microcontrollers & nRF52840 & MIMXRT1062 & STM32L4S5 &STM32F767\\
\hline
Processor& ARM Cortex-M4 &ARM Cortex-M7 & ARM Cortex-M4&ARM Cortex-M7\\
Clock (Mhz)& 64  & 600 & 120 &216\\
Flash (Mbytes) &  1 & 8 & 2 & 2 \\
SRAM (Kbytes)&  256 & 1000 &640 & 512\\
Wireless connectivity & yes & no & yes (external) & no\\
\hline
\end{tabular}
\end{table*}
The microcontroller is an essential component of an edge computing system. 
The primary functions of the MCU in this work are the following three aspects: sensor data acquisition, data storage, and real-time context extraction.  
\cref{tab: microcontroller} presented some critical parameters of four MCUs with ARM Cortex-M series processor cores and with different memory (RAM) and storage (Flash) sizes as well as clock speeds. 
In DL-based models, SRAM size constrains the activation size (read and write) and Flash constrains the model size (read-only) \cite{lin2020mcunet}. 
Moreover, the inference time is determined by the neural network model and the MCU's clock speed. 
Therefore, the MCU nRF52840 with ARM Cortex-M4 core from Arduino Nano Sense board and the high-performance MCU MIMXRT1062 with ARM Cortex-M7 from the Teensy 4.1 board were selected for this work.  
The system clock of MIMXRT1062 is up to 600 MHz, by which application scenarios with high real-time requirements could be met at the expense of higher power consumption than nRF52840 MCU. 
The MCU nRF52840 has a slower system clock (64 MHz) and an on-chip Bluetooth module providing a wireless interface by which the extracted context and label can be transmitted.
Besides, to investigate the performance of the tiny machine learning model for kitchen activity recognition, these models were also deployed on another two MCUs with similar processor cores (STM32L4S5 and STM32F767). 
All of them are supported by the TensorFlow Lite framework for tiny machine learning. 

\begin{figure*}[!t]
     \centering
     \begin{subfigure}[b]{0.24\textwidth}
         \centering
         \includegraphics[width=\textwidth]{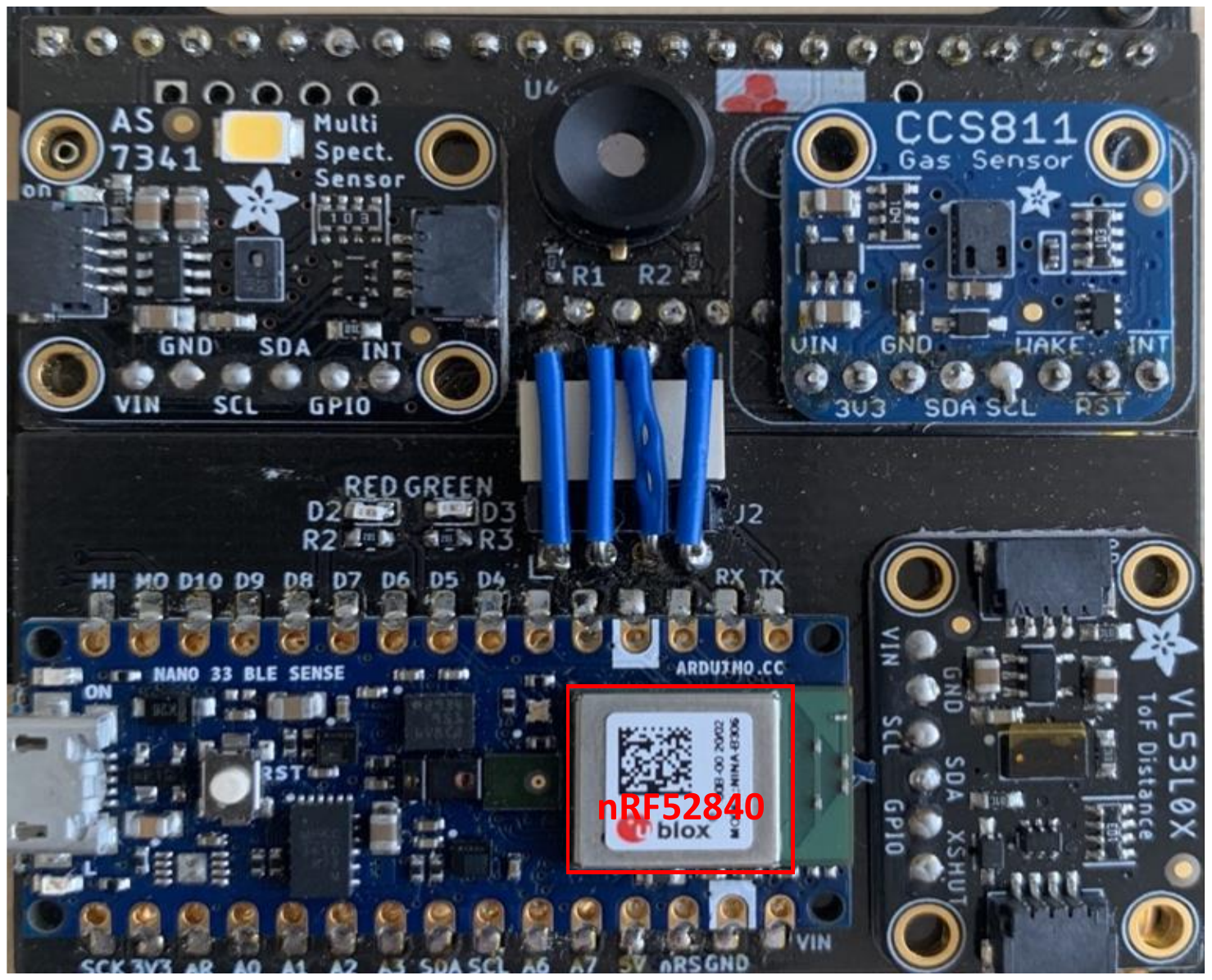}
         \caption{nRF52840}
         \label{fig:nRF52840}
     \end{subfigure}
     \hfill
     \begin{subfigure}[b]{0.24\textwidth}
         \centering
         \includegraphics[width=\textwidth]{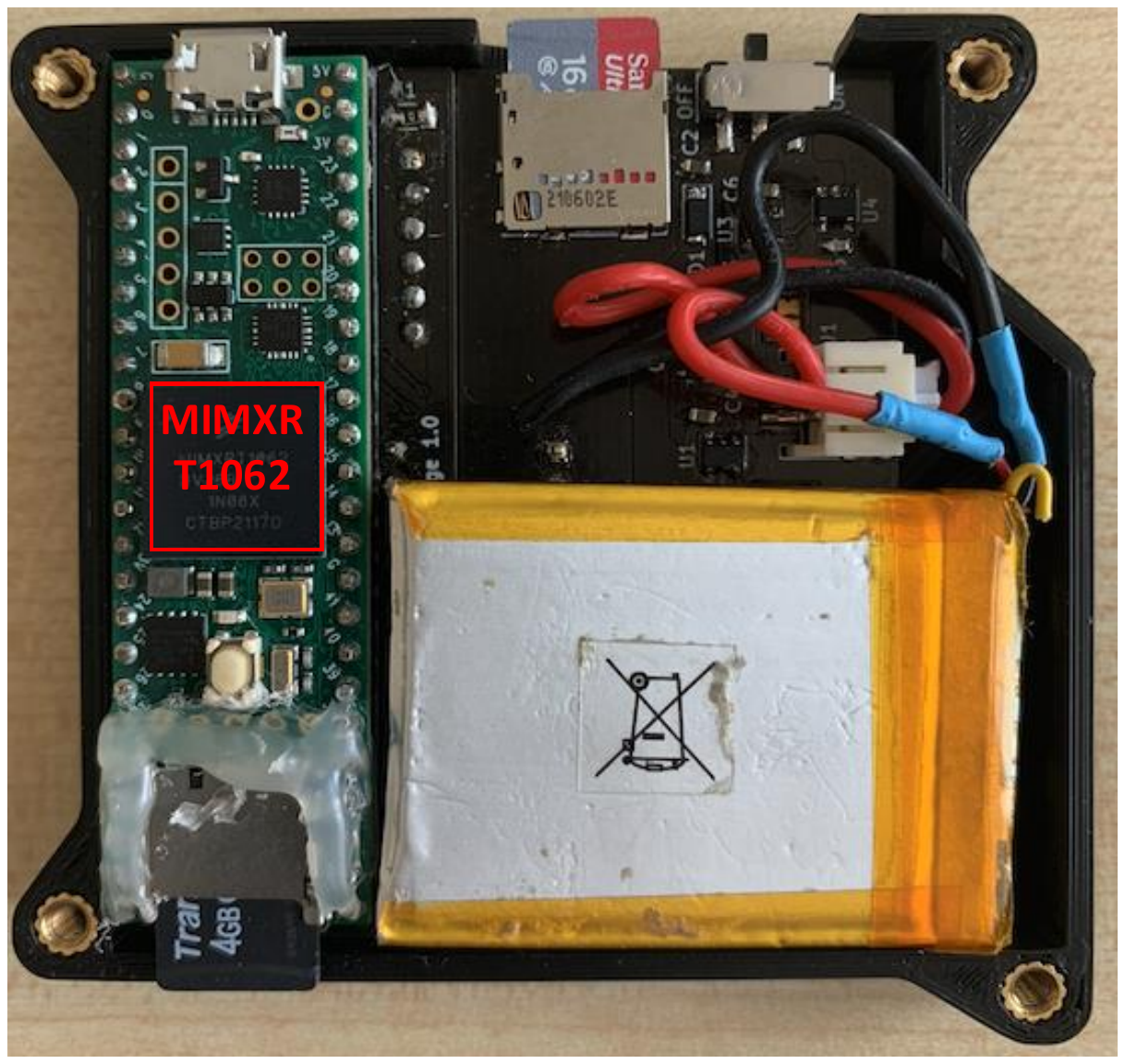}
         \caption{MIMXRT1062}
         \label{fig:MIMXRT1062}
     \end{subfigure}
     \hfill
     \begin{subfigure}[b]{0.24\textwidth}
         \centering
         \includegraphics[width=\textwidth]{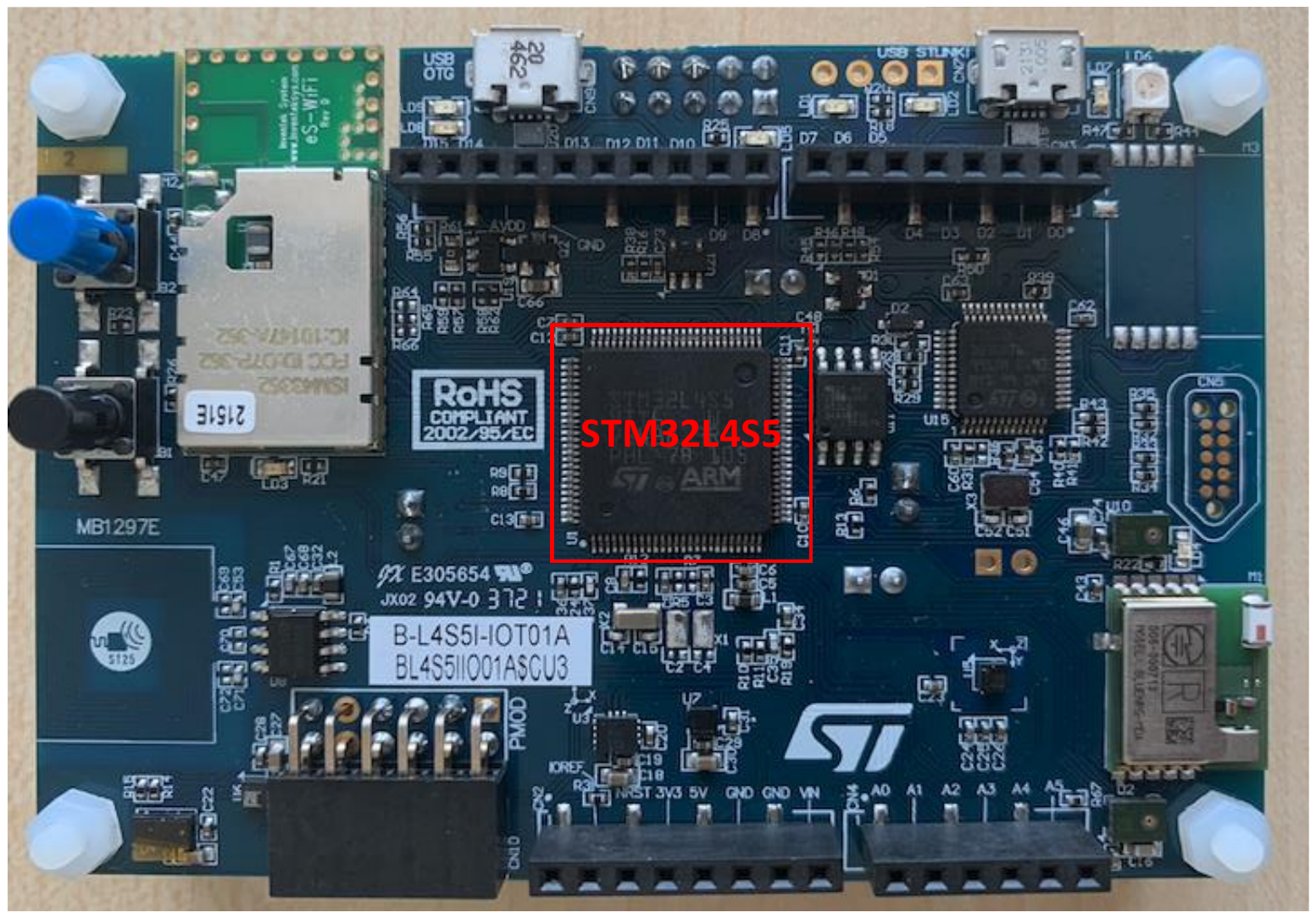}
         \caption{STM32L4S5}
         \label{fig:STM32L4S5}
     \end{subfigure}
     \hfill
     \begin{subfigure}[b]{0.24\textwidth}
         \centering
         \includegraphics[width=\textwidth]{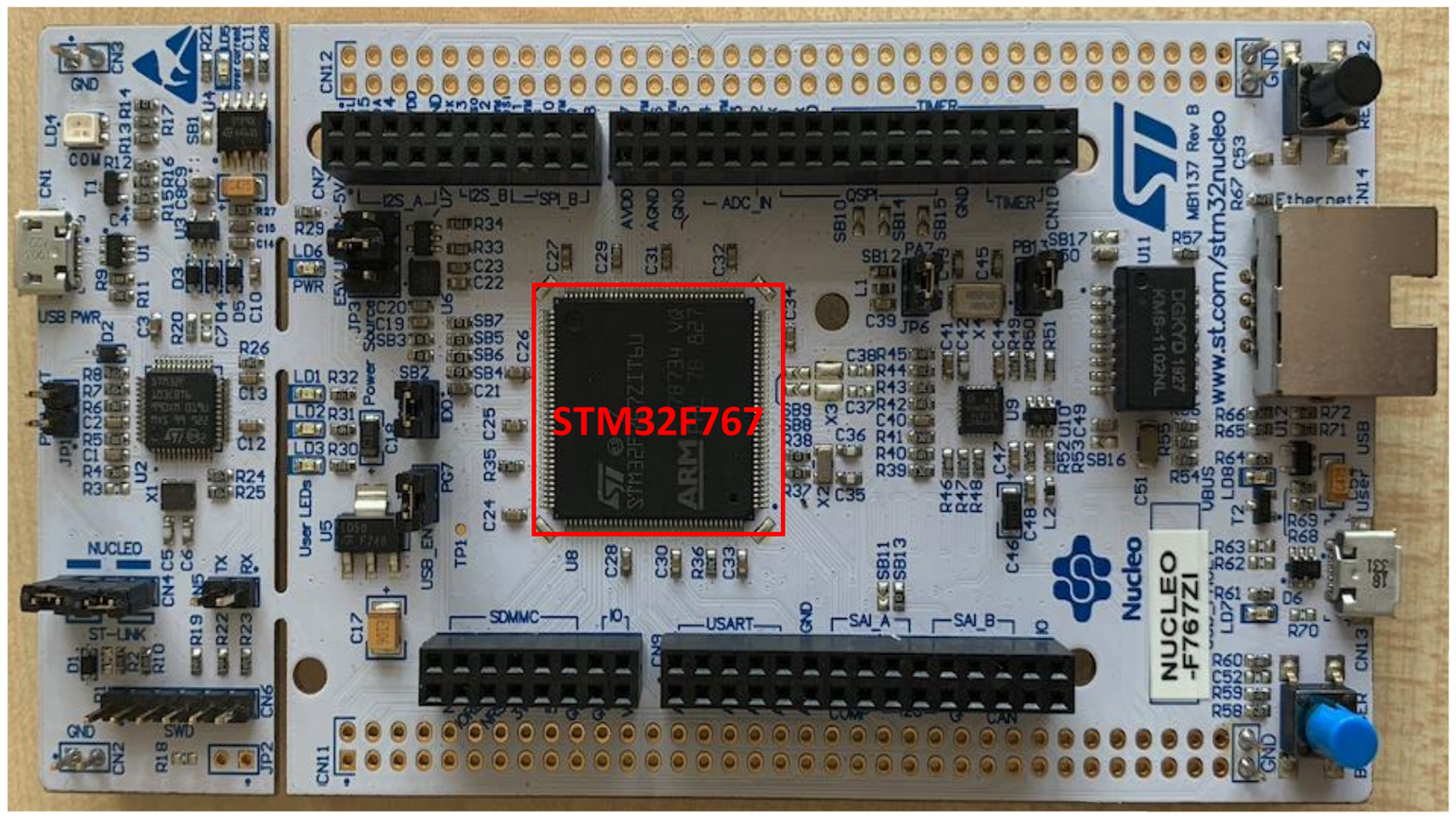}
         \caption{STM32F767}
         \label{fig:STM32F767}
     \end{subfigure}
    \caption{Hardware Prototype and Commercial Board. (a) the front view and (b) the back view of the prototype of the proposed edge-computing device; (c) NUCLEO -767ZI board; (d) STM32L4S5 Discovery kit}
    \label{fig:microcontroller}
\end{figure*}

\subsection{Neural Network Architecture}

Seven hundred ninety-one channels data from six sensors were recorded, and two kinds of information fusion methods based on the neural network were utilized in our previous work \cite{liu2022smart}. 
The feature fusion method with 791 channels of input data has achieved the highest recognition accuracy and macro F1-Score, however, its size is the largest.
The data fusion method concatenating all the channel data before inputting to the neural network has the most straightforward architecture, and the best result of the data fusion method is very close to the feature fusion method.
As the memory size is a limited resource of MCUs, the data fusion concatenating channels as input were selected to build a neural network running on the MCU in this work. Besides, one CNN layer was removed to reduce the model size compared to our previous work.

\begin{figure*}[!t]
\footnotesize
\centering
\includegraphics[width=0.8\linewidth]{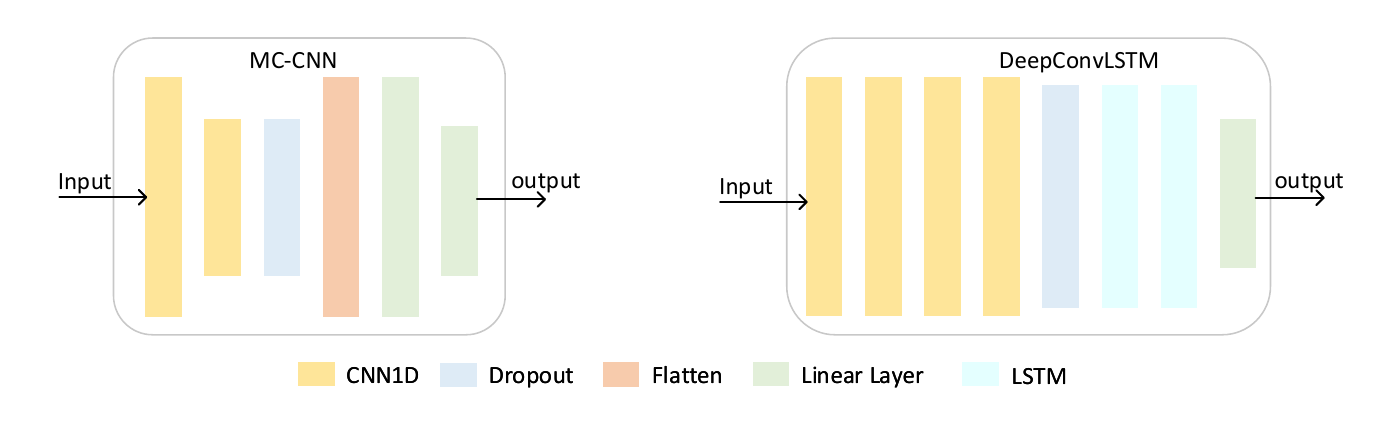}
\caption{Implemented Neural Network Architecture}
\label{fig: neural_network}
\end{figure*}

\cref{fig: neural_network} shows the architecture of the neural network utilized in this work. 
At first, the selected channel data from sensors were concatenated directly. 
Then, the concatenated data were fed into two 1D-CNN layers with a ReLu activation function, a dropout, and 1D average pooling layers to extract features. 
At last, the extracted features were input into dense layers to get an activity recognition result. 
The first layer of the 1D CNN has four times the number of filters as the second 1D CNN layer. 
In order to achieve the best activity recognition results while keeping the model size small,
the effect of hyperparameters, like filter number in CNN and input channel, on recognition results and model size was investigated.
Theoretically, the number of filters of the 1D CNN has an essential effect on recognition results.
Too many filters in CNN could result in over-fitting and a more extensive model size, however, fewer filters make the model size small and suitable for tiny machine learning, but it often causes under-fitting. 
Three filter size numbers of the first 1D CNN layer were selected: 128, 256, and 400 filters.
The input data were also divided into four groups in terms of sensor features with different input channels, such as 791 channels (including all sensor data), 768 channels (only infrared array sensor data), 23 channels (including all sensor data except infrared array sensor data), 17 channels (including all sensor data except infrared array sensor data, accelerometer, and gyroscope data).  

The neural network model utilized for kitchen activity recognition was built under \textbf{TensorFlow 2.10.0} framework and the model training process was performed on a laptop with the GeForce RTX 3080 Ti GPU. 
Then, the TensorFlow Lite model was generated by the TensorFlow Lite Converter. 
During conversion, optimization mechanisms like quantization, pruning, and clustering can be applied to reduce the model size and latency with minimal or no loss in accuracy. 
Here, the full integer quantization method was used, by which the latency and peak memory usage can be reduced.
All model math is integer quantized so that this model can be executed on the hardware only supporting integer operation. It is worth noting that the range, i.e., (min, max) of all floating-point tensors in the model needs to be estimated. Therefore, a representative dataset is required by the converter.
The training dataset was used as a representative dataset in this work. 

To evaluate the performance of real-time recognition performed on MCU, two models with different precision representations were generated for this task as follows:

\begin{enumerate}
    \item TFLite Model: converted TensorFlow Lite model without any optimization from TensorFlow Model by TensorFlow Lite converter, the model is represented with 32-bit precision float data
    \item Full integer Model: converted TensorFlow Lite model with full integer quantization from TensorFlow Model by TensorFlow Lite converter, the model is represented with 8-bit precision integer data.
\end{enumerate}

Since human activities are made of complex sequences of motor movements, and capturing these temporal dynamics is fundamental for successful HAR \cite{ordonez2016deep}, a DeepConvLSTM model, which includes four convolutional layers, two LSTM layers, and one softmax layer, was also utilized to recognize these kitchen activities and deployed on the microcontrollers. A performance comparison for this task between DeepConvLSTM model and MC-CNN model was given. 


\section{Evaluation of Kitchen Activity Recognition in Real-Time}
\label{sec:Evaluation}
\subsection{Dataset of Kitchen Activity}
\begin{figure*}[!t]
\footnotesize
\centering
\includegraphics[width=1.0\linewidth]{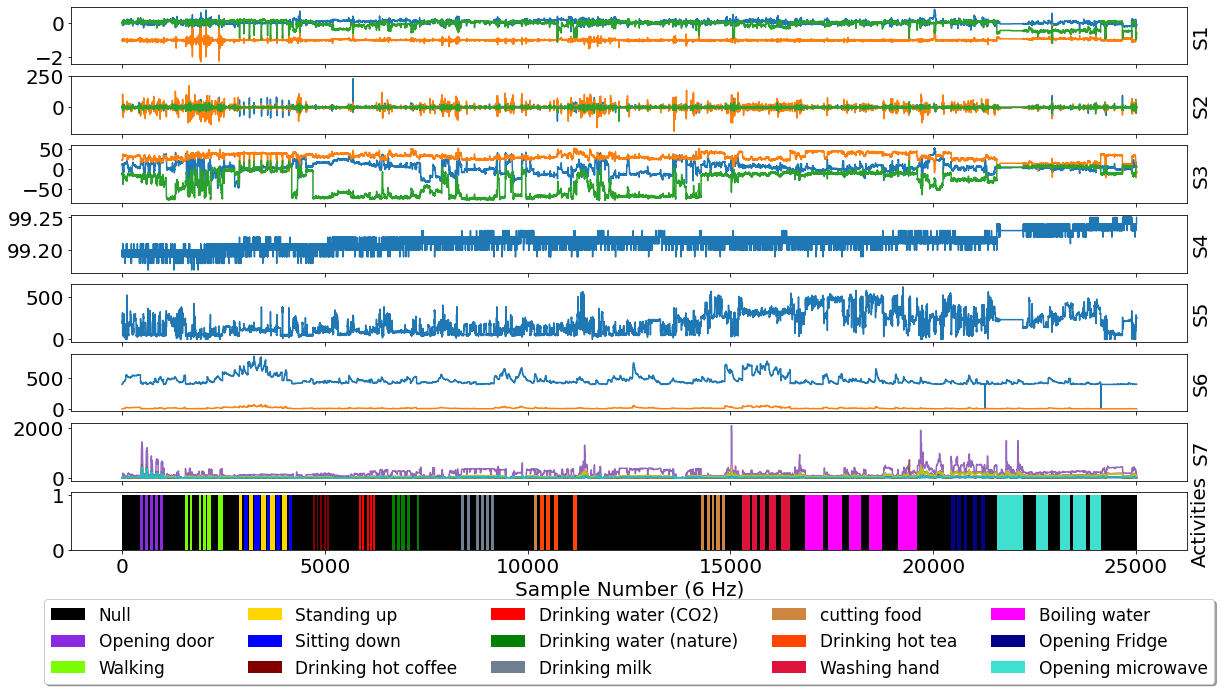}
\caption{Example of the measured multi-modal raw data from one volunteer and activity label. (\textbf{S1}: 3-channel accelerometer signal; \textbf{S2}: 3-channel gyroscope signal; \textbf{S3}: 3-channel magnetometer signal; \textbf{S4}: 1-channel Barometer signal; \textbf{S5}: 1-channel Distance signal; \textbf{S6}: 2-channel CO2 and TVOC gas signal; \textbf{S7}: 10-channel Optical Signal)}
\label{fig: raw_data}
\end{figure*}

The dataset used to train this work's neural network is from previous work\cite{liu2022smart}, 15 kinds of activity including the null class in the kitchen scenarios shown in \cref{fig: raw_data} were selected to recognize. 
Ten volunteers performed them in a realistic unmodified kitchen environment. 
The experiment was divided into five sessions, and the experiment lasted around one hour per volunteer. 
Most of the activities in this experiment strongly correlate with users' eating habits, e.g., recognizing beverage intake during consumption and in the case of food preparation, such as opening the microwave and boiling water. 

\subsection{Metrics of Comparison}
To assess the performance of our proposed device in a kitchen environment, we chose five metrics: accuracy, macro F1 Score, model size, inference time, and power consumption. Given the varying durations of kitchen activities and a constant data slide window size, the dataset suffers from imbalance. As a result, overall classification accuracy does not serve as a suitable performance indicator. Thus, the macro F1-score was also adopted as a benchmark. The MCU possesses significantly less memory and storage compared to computers and smartphones, making model size a vital metric for tiny machine learning applications. The main goal for deploying a machine model on an MCU is to facilitate local, real-time inferences. Nevertheless, due to constraints in computational resources such as the floating point unit (FPU) and clock speed, inference times can differ across MCUs. Additionally, power consumption is a also crucial metric for wearable devices as it affects battery life. Consequently, these five metrics were chosen as the evaluation criteria.

\subsection{Result of Recognition Accuracy, Macro F1 Score and Model Size}

\cref{tab: acuracy_result} shows the recognition results from two types of models with varying hyperparameters, such as the number of input features and the count of filters in the first layer of the 1D CNN. In the MC-CNN architecture, the initial CNN layer contains four times as many filters as its subsequent layer. In contrast, in the DeepConvLSTM architecture, the filter count remains consistent across all four layers. The highest accuracy achieved was 87.93\% by the TFLite model with the MC-CNN architecture, configured with 23 input channels and 400 filters in its initial layer. The same configuration in the DeepConvLSTM architecture reached the best accuracy of 84.01\%. Conversely, the lowest accuracy was observed in the Full Quantization Model using MC-CNN architecture with only 768 channel infrared sensor data, a similar result can be observed in the model with the DeepConvLSTM architecture. Moreover, changing the number of filters in the MC-CNN models had minimal impact on accuracy, improving it by merely one percent when filters were increased from 128 to 400. However, in DeepConvLSTM models, increasing filters from 32 to 100, accuracy was improved by over three percent. Despite significant reductions in size, the Full Quantization Model maintained nearly the same accuracy and macro F1 score as the TFLite Model.

The F1 score, which assesses model performance by balancing precision and recall, indicates that input sensor modalities greatly influence the macro F1 score; it rose by about ten percent with an increase in sensor types. The TFLite MC-CNN model achieved the highest macro F1 score of 77.96\%. Given the variety of activities ranging from movement to drinking. While more sensor modalities and a greater number of convolutional filters can enhance feature extraction, improving the macro F1 score, it leads to the increase of the model size.

\begin{table*}[!t]
\centering
\renewcommand{\arraystretch}{1.0}
\footnotesize
\caption{Result Summery of Recognition Accuracy and Macro F1 Score 
(MC-CNN: N1=128, N2=256, N3=400. DeepConvLSTM: N1=32, N2=64, N3=100.  The best accuracy results are denoted in red color, and the poorest results are denoted in bold font)}
\label{tab: acuracy_result}
\begin{tabular}{ c c c|p{1.5cm} p{1.5cm} p{1.5cm}|p{1.5cm} p{1.5cm} p{1.5cm}}
\hline
\multicolumn{3}{c|}{Parameters}&\multicolumn{3}{c|}{MC-CNN}&\multicolumn{3}{c}{DeepConvLSTM}\\
\hline
Channels & Filters  &Inference Model & Macro \newline F1 Score  & Accuracy& Model Size (KBytes)&Macro \newline F1 Score & Accuracy& Model Size (KBytes)\\
\hline

\multirow{6}{*}{17} & \multirow{2}{*}{N1} & TFLite Model & 67.94 & 84.31 & 107.51 & 62.64 & 77.84 & 935.38\\ 
~ & ~ & Full Quantization Model & \textbf{67.89} & 84.36 & \textbf{36.91} & \textbf{61.81} & 77.27 & \textbf{259.72}\\ \cline{2-9}

~ & \multirow{2}{*}{N2} & TFLite Model & 70.07 & 84.97 & 299.12 & 67.16 & 81.31 & 1118.55\\ 
~ & ~ & Full Quantization Model & 69.93 & 84.82 & 89.14 & 66.88 & 80.98 & 308.97\\\cline{2-9} 

~ & \multirow{2}{*}{N3} & TFLite Model & 72.33 & 85.26 & 632.20 & 67.83 & 82.06 & 1412.74\\ 
~ & ~ & Full Quantization Model & 72.00 & 85.13 & 177.26 & 67.29 & 81.59 & 386.40\\ \hline

\multirow{6}{*}{23} & \multirow{2}{*}{N1} & TFLite Model & 72.53 & 86.56 & 116.72 & 65.65 & 79.96 & 937.69\\  
~ & ~ & Full Quantization Model & 72.25 & 86.39 & 39.22 & 64.50 & 79.43 & 260.29\\ \cline{2-9}

~ & \multirow{2}{*}{N2} & TFLite Model & 75.19 & 87.26 & 317.56 & 70.94 & 83.80 & 1123.16\\ 
~ & ~ & Full Quantization Model &74.97 & 87.16 & 93.74 & 69.16 & 83.10 & 310.12\\ \cline{2-9}

~ & \multirow{2}{*}{N3} & TFLite Model & 77.25 & \textcolor{red}{87.93} & 661.00 & 72.10 & \textcolor{red}{84.01} & 1419.94\\ 
~ & ~ & Full Quantization Model & 77.05 & 87.83 & 184.46 & 71.12 & 83.56 & 388.20\\ \hline

\multirow{6}{*}{768} & \multirow{2}{*}{N1}& TFLite Model & 71.95 & 82.24 & 1261.04 & 62.84 & 75.63 & 1223.77\\ 
~ & ~ & Full Quantization Model & 71.27 & 82.20 & 325.30 & 62.55 & \textbf{75.28} & 331.81\\ \cline{2-9}

~ & \multirow{2}{*}{N2} & TFLite Model & 71.56 & \textbf{81.91} & 2606.20 & 66.37 & 77.26 & 1695.32\\ 
~ & ~ & Full Quantization Model & 71.74 & 82.30 & 665.90 & 64.49 & 76.23 & 453.16\\ \cline{2-9}

~ & \multirow{2}{*}{N3} & TFLite Model & 71.60 & 82.03 & 4237.00 & 68.29 & 79.35 & 2313.94\\ 
~ & ~ & Full Quantization Model & 71.59 & 82.54 & 1078.46 & 66.92 & 78.53 & 611.70\\ \hline

\multirow{6}{*}{791} & \multirow{2}{*}{N1} & TFLite Model & 76.67 & 86.45 & 1296.37 & 67.51 & 78.88 & 1232.60\\ 
~ & ~ & Full Quantization Model & 76.53 & 86.35 & 334.13 & 66.46 & 78.23 & 334.02\\ \cline{2-9}

~ & \multirow{2}{*}{N2} & TFLite Model & 77.68 & 86.75 & 2676.85 & 71.59 & 82.23 & 1712.98\\ 
~ & ~ & Full Quantization Model & 77.61 & 86.92 & 683.57 & 70.95 & 81.71 & 457.57\\ \cline{2-9}

~ & \multirow{2}{*}{N3} &TFLite Model & \textcolor{red}{77.96} & 87.05 & \textcolor{red}{4347.40} & 72.48 & 82.81 & \textcolor{red}{2341.54}\\ 
~ & ~ & Full Quantization Model & 77.84 & 87.10 & 1106.06 & \textcolor{red}{72.55} & 82.61 & 618.60\\ \hline

\hline
\end{tabular}
\vspace{-5mm}
\end{table*}

According to \cref{tab: acuracy_result}, the TFLite Model is approximately four times larger than the Full Quantization Model due to its use of 8-bit integers instead of 32-bit float data. Among all models, the largest is the TFLite MC-CNN architecture at 4346 KBytes with 791 input channels, while the smallest is a Full Quantization Model at only 36.91 KBytes with 17 input channels. The presence of fully connected layers in the MC-CNN architecture causes a significant increase in model size with the number of input channels, though this effect is less pronounced in the DeepConvLSTM architecture. The smallest model size for DeepConvLSTM is about four times that of MC-CNN's smallest, but its largest model size is only half that of MC-CNN's largest. Enhancements in filters and input numbers improve recognition performance but also significantly increase the model size from kilobytes to megabytes. However, given that most MCUs, as shown in \cref{tab: microcontroller}, have limited flash and SRAM sizes of up to 2 MBytes and 1 MByte respectively, many larger TFLite models may not be suitable. The largest Full Quantization Model size is around 1.1 MBytes, making full integer quantization a practical optimization strategy for these models, trading off minimal accuracy loss for substantial size reduction.

\subsection{Result of Model Inference on Edge devices}

\begin{table*}[!t]
\centering
\renewcommand{\arraystretch}{1.1}
\footnotesize
\caption{Result of Inference Time and Power Consumption(the best accuracy results are denoted in red color, and the poorest results are denoted in bold font)}
\label{tab: inference time}
\begin{tabular}{ c c c c c c c}
\hline
Architecture & Metrics & Inference Model &\textbf{nRF52840} & \textbf{MIMXRT1062} & STM32L4S5 & STM32F767\\
\hline
\multirow{4}{*}{MC-CNN} & \multirow{2}{*}{Inference Time (ms)} & TFLite Model& -&169.63 & \textbf{4540.4}& 575.80\\
 & &Full Quantization Model &394.49 & \textcolor{red}{25.26}&195.21& 31.57\\
\cline{2-7}
& \multirow{2}{*}{Power Consumption (W)} & TFLite Model & - & 0.78 &0.67 &1.13 \\
& &Full Quantization Model &\textcolor{red}{0.10} & 0.73 & 0.62 & 1.08\\
\cline{2-7}

\hline
\multirow{4}{*}{DeepConvLSTM} & \multirow{2}{*}{Inference Time (ms)} & TFLite Model&-&253.84 & 2561.40&  346.48\\
 & &Full Quantization Model & 685.95 &56.04 &516.07& 70.15 \\
\cline{2-7}
& \multirow{2}{*}{Power Consumption (W)} & TFLite Model&- & 0.72&0.68 & 1.13\\
& &Full Quantization Model & \textcolor{red}{0.10}& 0.77& 0.79& \textbf{1.14} \\
\cline{2-7}

\hline

\end{tabular}
\end{table*}

To assess the real-time performance and power consumption of the model inference on the designated hardware, we opted for models with 23 input channels for implementation on microcontrollers due to their superior accuracy with a model size of approximately 1 Mbytes. Additionally, two other MCUs from development boards with comparable processors, as indicated in \cref{tab: microcontroller}, were utilized to benchmark the real-time performance against our proposed hardware system. The primary testing sequence comprised reading sensor data, starting the timer counter, performing inference, predicting activities, measuring inference time, and handling outputs. Given that the six sensors for kitchen activity recognition were absent on the development boards, the sensor data, in 32-bit float format, was stored in these MCUs to evaluate model inference performance. Should the test model be a Full Quantization Model, the sensor data would be converted to an 8-bit integer prior to being fed into the model. A timer with microsecond accuracy was employed to record the inference time. The output is an array of 15 elements, each representing the probability of a specific activity. The activity predictor identifies and selects the element with the highest probability as the final output. Subsequently, the output handler activates the serial port to send the recognition result and inference time to the laptop. To measure the power consumption during inference, an Oscilloscope was used along with a one Ohm Resistor placed in series in the 5V power supply line, with oscilloscope probes attached to both terminals of the resistor, allowing the inference current to be measured. The power consumption is determined by multiplying the measured current by the voltage.

\cref{tab: inference time} illustrates the hardware performance of selected models on four distinct microcontrollers. Given that nRF52840 possesses only 256 Kbytes of SRAM and 1 Mbyte of storage, it cannot support the TFLite Model due to its size. Thus, only the Full Quantization Model was tested on nRF52840. The fastest inference time was 25.26 milliseconds on the MIMXRT1062, a high-performance microcontroller with the fastest system clock. Conversely, the slowest inference time of 4540.4 milliseconds was realized on the STM32L4S5, which was running the TFLite Model with the MC-CNN architecture. Additionally, the inference time for the Full Quantization Model is significantly shorter than that of the TFLite Model. For instance, running the TFLite Model with MC-CNN architecture on STM32F767 takes about 18 times longer than the Full Quantization Model, primarily because floating-point operations consume much more time than 8-bit precision integer operations. The lowest power consumption during inference, at just 0.1 W, was achieved by nRF52840, whereas the highest was 1.14 W by STM32F767. It is also noted that model quantization has minimal impact on power consumption in this study.

\section{Discussion}
\label{sec: disccussion}
As indicated by the results in \cref{tab: acuracy_result}, the presence of additional sensor modalities and convolutional layer filters enhances the accuracy of kitchen activity recognition, though their impact is minimal when compared to the increase in model size and the decrease in inference time. In practical settings, a neural network model operating on an edge device and processing data locally offers numerous benefits, such as enhanced privacy protection. Therefore, considerations of model size and inference time become essential for effective deployment on edge devices. In the context of our application, the MC-CNN architecture outperforms DeepConvLSTM in terms of recognition accuracy and inference time, the former also is approximately half the size and faster in inference than the latter. Consequently, the MC-CNN architecture is more suitable for our purposes. Theoretically, optimization methods for neural networks do influence power consumption, but this impact is negligible when comparing the power usage among four microcontrollers due to other dominant factors such as system clock speed and hardware architecture. In our experiments, the microcontroller MIMXRT1062, with the highest clock speed of 600 MHz, demonstrated superior real-time inference capabilities. In contrast, the microcontroller nRF52840 exhibited the lowest power consumption of the group. Thus, our proposed Wearable Multi-Modal Edge-Computing System is adaptable for various real-time HAR scenarios, allowing for the selection of a microcontroller based on desired performance or energy efficiency.

\section{Conclusion}
\label{sec: conclusion}
In this paper, we introduced a wearable multi-modal edge-computing system designed for real-time kitchen activity recognition. Initially, we explored how sensor channel inputs and convolutional layer filters impact both recognition accuracy and model size, examining the performance of 48 TensorFlow Lite models across various input channels, filter sizes, optimization techniques, and neural network architectures. The highest achieved recognition accuracy was 87.93\% with a macro F1 score of 77.25\%. Subsequently, the most accurate model was implemented on four microcontrollers to assess its real-time performance and energy efficiency. The MIMXRT1062 microcontroller demonstrated a fast inference time of 25.26 milliseconds for recognizing 15 types of activities, while the nRF52840 microcontroller exhibited the lowest energy consumption at 100 mW. These results demonstrate that the developed system holds significant promise for real-time HAR. Future work will focus on refining the tiny machine learning model and broadening its application contexts. 
%
%
%
\bibliographystyle{splncs04}
\bibliography{refs}
%





\end{document}